\documentstyle[12pt,epsf]{article}
\begin{document}
\setlength{\baselineskip}{0.3in}


\newcommand{\beq}{\begin{equation}}
\newcommand{\eeq}{\end{equation}}
\newcommand{\beqa}{\begin{eqnarray}}
\newcommand{\eeqa}{\end{eqnarray}}
\newcommand{\lsim}{\begin{array}{c}\,\sim\vspace{-21pt}\\<  
\end{array}}
\newcommand{\gsim}{\begin{array}{c}\sim\vspace{-21pt}\\> \end{array}}
\newcommand{\nt}{\nu_\tau}
\newcommand{\nee}{\nu_e}
\newcommand{\nm}{\nu_\mu}
\newcommand{\bi}{\bibitem}

\begin{titlepage}
\pagestyle{empty}
\rightline{UCSD-TH-96-09}
\rightline{hep-ph/9605260}
\rightline{May 1996; Revised: June 1996 }
\vspace*{0.5cm}
\begin{center}
\vglue .06in
{\Large \bf  ~Errors in Lattice Extractions
of $\alpha_s$ Due to Use of Unphysical Pion Masses 
} \\[0.8in]
{\sc Benjamin Grinstein and  I.Z.~Rothstein  }
\vskip 0.3cm
\begin{tabular}{c}
{\it Dept.  of Physics}\\
{\it UCSD}\\
{\it La Jolla Ca. 92093-0319 }\\[.15in]
\end{tabular}
\vskip 0.25cm
\vskip 1truecm

\vspace*{0.5cm}

{\bf Abstract}\\[-.05in]

\begin{quote}
We investigate the errors due to the use of unphysical values 
of light quark
 masses in lattice extractions of $\alpha_s$. 
A functional form for the pion mass dependence of the quarkonium
mass splittings ($\Delta m$) 
is given as an expansion in $m_\pi/(4\pi f_\pi)$ and $m_\pi
r_B$, where $r_B$ is the quarkonium Bohr radius. We find that, to lowest order,
 $\Delta m\simeq A+B m_\pi^2$, where the scale of $B$ is given
by $f_\pi^2 r_B^3$.
To order $m_\pi^4$ there are four unknown coefficients, however, 
utilizing  multipole and operator product expansions, 
symmetry arguments  eliminate one of  the four
unknowns. 
Using the central values for the lattice spacings which were
extracted using two different, unphysical values for the pion mass, 
we find that the
errors introduced by extrapolating to the physical regime are
comparable to the errors quoted due to other sources.  
Extrapolation to physical values of the pion mass {\it increases}
the value of $\alpha_s(M_Z)$,  bringing its value closer to the 
high energy extractions.

\end{quote}
\end{center}

\end{titlepage}
\setcounter{footnote}{0}

\newpage

\section{Introduction}
\indent
Precision measurements of $\alpha_s$ are motivated  not only by our need 
for more precise predictions within the confines of perturbative QCD,
but also by the possibility that such a measurement could lead
to hints of new physics. For instance, GUTS often make predictions for
the value of $\alpha_s$ at low energies, and thus, these theories
may be tested via a precise determination of $\alpha_s$.
Presently, there seems to be a slight discrepancy  between low and high energy
extractions of $\alpha_s$ \cite{alt95}, and it has become
crucial to have a good handle on the errors involved before we
can determine whether or not this discrepancy is a signal for new
physics.
One of the low energy extractions, and the
one with the smallest quoted error bar, is performed via a  measurement of  the
splittings between quarkonia levels on the lattice.

There have been several lattice extractions of $\alpha_s(M_Z)$\cite{NRQCD1}
\cite{others}, with the smallest quoted error being given in \cite{NRQCD1}. 
The  extractions are performed by determining
the lattice spacing via a  measurement of  the splittings between different 
quarkonia
energy levels, the values of which are known from laboratory experiments.
Unfortunately, due to computational difficulties,
 the measurements are performed in a theory with 
 unphysical values for the masses of the light mesons.
The values of $\alpha_s$ extracted in these unphysical
theories are then assumed to differ little from the physical
values. The justification for this assumption is that as long
as the light quark masses are small compared the typical
momenta in the bound state, the level splittings should
be insensitive to the quark masses.

In this note we will determine the size of these errors in a
systematic expansion in the pion mass.  More specifically, we address
the errors incurred by extrapolation to physical values of the light
quark masses and the assumption of $SU(3)$ flavor symmetry.  We use
previously developed techniques \cite{vol79}\cite{pes79}\cite{bha79}
to determine an appropriate effective field
theory to describe the interaction of light mesons with
heavy $Q\bar{Q}$ bound states. Once the correct effective field
theory is found we are able to calculate the errors introduced
due to extrapolation to physical values with no guess work
on unknown functional forms.

The remainder of this paper is structured as follows. In the next
section we will review the lattice measurements with emphasis on the
approximations used. In section three we present our results of the
errors.  In section four we write down the chiral Lagrangian which
describes the interaction of the quarkonia with the light degrees of
freedom.  In section five we show that, within the confines of a
multipole expansion, it is possible to map the theory of heavy quarks
onto a chiral Lagrangian using symmetry arguments first noted in
\cite{vol80}.  The final section is reserved for conclusions and the
need for further computations.

\section{Present Lattice Extractions}
\indent
The accuracy of present lattice calculations are limited by the inability
to properly simulate the light quarks. In general, it is difficult
enough to include dynamical fermions in the calculations, no less
calculate using their physical masses. Thus, to measure physical
parameters with accuracies which are competitive with experiment,
it is necessary to look for observables whose values are
not  strongly dependent on the dynamics of the light quarks.
This is one of the primary reasons why heavy quark anti-quark bound states 
have been chosen as laboratories for measuring the value of $\alpha_s$.
The level splittings in quarkonia are  expected to be
insensitive to the masses and the number of light quarks due to the
fact that the Bohr radius is so small compared to the 
hadronic scale $1/\Lambda_{QCD}$.
It is therefore hoped that calculating with unphysical masses for
the light mesons should be a very good approximation, and the
computational problems which arise in the physical region of
parameter space can be avoided.

In this section we review the method by which the measurements are
made following the work of ref \cite{NRQCD1}. There are several
sources of errors in these calculations, all of which are claimed to
be under control. Here we will focus only on those errors induced by
extrapolation to physical values of the light quark masses.

The coupling is measured by first determining the lattice spacing, $a$,
 in a theory with some unphysical value for the pion and kaon masses.
This is done by measuring the level splittings in
quarkonia and then extracting the value of $a$ by assuming that 
the value of the splitting in this  unphysical theory is very
close to physical value.  
Once the lattice spacing is known, the coupling $\alpha_P$ is
determined by its plaquette value. 
\beq  
-\ln W_{1,1}=\frac{4\pi}{3}\alpha_P(\frac{3.41}{a})\left[
1-(1.185+.070n_f)\alpha_P\right].
\eeq
The value of $\alpha_s$ in the $\overline{MS}$ is then determined through
the relation 
\beq
\label{ptoms}
\alpha_{\overline{MS}}^{n_f}(Q)=\alpha_P^{n_f}(e^{5/6}Q)\left[
1+\frac{2}{\pi}\alpha_P^{n_f}+O(\alpha_P^{n_f})^2\right]
.
\eeq

The measurements are performed with varying numbers (up to $n_f=2$) of
light quarks with several different, unphysical quark masses. The
latest published results from the NRQCD collaboration are shown in
table I\cite{NRQCD1}.  The physical values of $\alpha_s$ are then
found by extrapolating to $n_f=3$ using the fact that the inverse
coupling $1/\alpha_P^{(n_f)}$ is known to be almost linear for small
changes in $n_f$.   The
assumption is made that the results are independent of the pion
mass  and  $SU(3)$ symmetry breaking effects are ignored.  No errors 
are quoted for these parts of the
procedure.

\renewcommand{\thetable}{\Roman{table}}
\begin{table}
\centering
\caption{Measurements of $\alpha_P(8.2~GeV)$.}
\begin{tabular}{||l|r|r||} \hline
                  & 1S-2S & 1-S-1P \\ \hline
$am_q=.01,~n_f=2,~am_\pi=0.419(2)$      &  0.1793(16)        &     0.1777(23)  
\\ \hline
$am_q=.025,~n_f=2,~am_\pi=0.269(1)$        &    0.1760(35) &        0.1735(28) 
\\ \hline
$n_f=0$            & 0.1551(11)& 0.1505(9) \\ \hline\hline
\end{tabular}
\end{table}
The value of the coupling at the scale $M_Z$ is then determined by
first running down to the mass of the charm quark using the three flavor
beta function and then running back up to $M_Z$, taking
into account the $b$ quark threshold. Errors due to variations
in the thresholds were shown to be negligible.
Using $m_qa=.01$ the values found were:
\begin{eqnarray}
\alpha_{\overline{MS}}(M_Z)&=&0.1152(24)~~~~~~~~~~~~1S-1P\\
\alpha_{\overline{MS}}(M_Z)&=&0.1154(26)~~~~~~~~~~~~1S-2S.
\end{eqnarray}
The errors quoted here were found assuming a coefficient of one for the
$\alpha_P^2$ term in (2).
\section{Results: Procedure for Calculating the Errors}
\indent
We will now present the results for the errors which will 
be derived in a later
section. In a QCD like theory with an arbitrary pion mass,
the dimensionless mass splitting can be written as
\begin{eqnarray}
\label{dimless}
a\Delta m&=&aA
+\frac{B}{a^3}f^2\left[\frac{1}{(4\pi f)^2}
\left((am_\eta)^4\log \frac{m_\eta^2}{4\pi\mu^2}+3(am_\pi)^4
\log\frac{m_\pi^2}
{4\pi \mu^2}+4(am_K)^4\log\frac{m_K^2}{4\pi \mu^2}\right)
\right. \nonumber
\\  
&-&
\left.
 6C((am_K)^2+\frac{(am_\pi)^2}{2})\right]
+\frac{D(\mu)}{a^316\pi^2}\left(3(am_\pi)^4+4(am_K)^4+(am_\eta)^4\right)
+....
\end{eqnarray}
Here $a$ is the lattice spacing, $A,B,C,D$ are dimensionful unknowns,
 and the higher order terms left out are 
suppressed by powers of $\frac{m_\pi^2}{4\pi f_\pi}$. 
We will show in a later section that within the confines of a 
multipole expansion, the coefficient $C$ is 1 at leading order under certain
conditions.
Furthermore, in general, it is be possible to calculate
$B$ using potential models, but
we will leave it as  an unknown for now.
The parameter $f$ is independent of $m_\pi$, and 
its value can be
calculated in terms of $m_\pi$ and $f_\pi$ in QCD. 
In a given theory with light  quark
mass $m_q$, we know only the dimensionless combination $am_\pi$.
In the two theories which have been investigated to
date $am_\pi=0.269(1),~0.419(2)$ for $am_q=0.01,0.025$ respectively.

Furthermore, all the simulations which have been performed 
with dynamical fermions have only included  two light quark flavors.
We therefore reduce the general form (\ref{dimless}) to the case of
$SU(2)$
\beq
\label{su2}
a\Delta m=\left[ a{A_2}+\frac{{B_2}}{a}f^2(am_\pi)^2\left(-2+
\frac{(am_\pi)^2}{(af)^2}\frac{1}{16\pi^2}\log \frac{m_\pi^2}{4\pi\mu^2}
\right)+\frac{{D_2}(\mu)}{16\pi^2a^3}(am_\pi)^4 \right].
\eeq
In the future,  when simulations are performed with three light quarks, 
we will be able to determine the corrections due to $SU(3)$ breaking using
(\ref{dimless}).
Notice that 
going to the $SU(2)$ case does not reduce the number of unknowns.

Since we have only two data points to fit, let us
 begin by considering the corrections only up
to $O(m_\pi^2)$.
In this case, we may solve for $A$ and $B$ and
subsequently determine the error induced by extrapolating
to physical masses for the pions. 
For the case of three flavors one should choose the value
$(m_\pi^2+2m_K^2)/3$ for the ``physical pion mass''. 
Given that in the extraction process we find $\alpha_P$ with
two flavors then extrapolate to three, we see no reason why this
should remain the correct value. Since we are calculating errors here, 
we believe that the proper choice should be somewhere between the 
$140$ and $410~MeV$.

Using the physical values
$\frac{\Delta m_{1S-1P}}{m_\pi}
=2.86$ and $\frac{\Delta m_{1S-2S}}{m_\pi}=4.00$ for the
$1S-1P$ and $2S-1S$ splittings in the Upsilon system respectively, we find 
for $m_\pi=140(410)~MeV$\begin{eqnarray}
a^{-1}\mid_{phys} &=&2.56~(2.47) ~~~~~~~~~~1S-1P\\
a^{-1}\mid_{phys} &=&2.53~(2.52) ~~~~~~~~~~1S-2S.
\end{eqnarray}
The authors of
\cite{NRQCD1} found  the value  $\alpha_P(8.32)=0.178$ $(1S-1P)$ and 
$\alpha_P(8.08)=0.179$ $(1S-2S)$ for $m_\pi=640$MeV, leading
to, up to quartic terms in the pion mass,
  $\alpha_P(8.73)=0.178$ $(1S-1P)$ 
and   $\alpha_P(8.63)=0.179$ $(1S-2S)$, 
when the pion mass takes on its physical value.
We have assumed here that  $lnW_{1,1}$ is independent of the pion
mass given that it is a short distance quantity.
Varying the pion mass was shown to change 
$\alpha_P$ by less than $0.2\%$ \cite{junko}.
Table II shows our results for the dependence of
$\alpha_{\over{MS}}(M_Z)$ on the value of the pion mass.

\renewcommand{\thetable}{\Roman{table}}
\begin{table}
\centering
\caption{Measurements of $\alpha_{\over{MS}}(91.2~GeV)$.}
\begin{tabular}{||l|r|r||} \hline
  $m_\pi~(MeV)$                & 1S-2S & 1S-1P \\ \hline
~~~910      &  0.1134        &     0.1149  
\\ \hline
~~~640        &   0.1154(26) &        0.1152(24) 
\\ \hline
~~~410        &0.1174        &    0.1167
\\ \hline
~~~140 &  0.1184    &0.1174
\\ \hline\hline
\end{tabular}
\end{table}

In calculating the size of the error here we must consider the fact
that we have assumed that the expansion (\ref{dimless}) is well behaved.
Indeed, in calculating the values of the coefficients we have used
the data for a theory in which the pion mass is on the order
$\sim900~MeV$, therefore these results should not be trusted at a quantitative
level but should be a qualitative estimate of the error.
To get an accurate value for the constants $A_2,B_2$ and 
$D_2$ it is important
that at least one other lattice simulation be performed with light quark
masses smaller than $am_q=0.025$.

We may make the guess that the non-analytic piece dominates 
the counter-term as is sometimes done when working
with chiral Lagrangians. Taking $4\pi \mu^2$ to be the lattice
spacing, we find that the results shift little. 
The net effect of the logs is to decrease the value of $\alpha_s (M_Z)$
at the level of $0.5\%$, which is to be compared to the $3\%$ effect 
found when just the piece quadratic in the pion mass is kept.

\section{Chiral Lagrangians and Heavy Quarks}
Chiral Lagrangians for heavy-light systems have been utilized
for several years \cite{bur92}.  For these systems there is only
one scale in the theory besides the hadronic scale, namely the mass of
the heavy quark.  This scale is trivially removed by rescaling the
mesonic field as was originally done in heavy quark effective field
theory.  One redefines the heavy
meson field according to
\beq
H(x)=e^{-im_Qv\cdot x}H^v(x)
\eeq
and treats this field as a classical static source labeled by its
velocity, $v$.  In so doing, spin symmetry as well as flavor symmetry
(between $D$ and $B$ mesons) becomes manifest\cite{wis93}. Once the
heavy quark mass has been scaled out, there is no question as to how
the operators scale.  

In the case of heavy-heavy systems, 
chiral symmetry has been utilized in decays for quite
sometime \cite{vol80}\cite{yan80}\cite{iof80}\cite{nov80}.
The chiral Lagrangian for these systems, while equivalent to the
current algebra approach utilized in the above references, organizes
the expansion perhaps more naturally.
Such Lagrangian 
 have only been written down
in the literature more recently \cite{cas93}\cite{man95}.
Though 
the heavy-heavy system does not posses a flavor symmetry, 
the spin symmetry  remains, contrary to what is claimed in
\cite{cas93}.
Spin symmetry breaking effects are  suppressed by powers
of the relative velocity of the heavy quarks  instead of inverse
powers of the mass as in the case of the heavy-light systems. 
The velocity scaling rules for the factorized
heavy quark matrix elements will be  the same as  those derived for the
NRQCD formalism \cite{lep92}. For our calculation, the spin symmetry
will not be relevant and therefore it will not be manifested in our
phenomenological Lagrangian.

Let us consider the chiral Lagrangian for the spin one $1S$ and $2S$ states. 
For the $1S$ state the lowest order Lagrangian,  
which is invariant under chiral symmetry is
given by
\begin{eqnarray}
\label{chiralL}
L_{int}&=&c_1 h^{(v)\mu}  h^{(v) *}_\mu  
 Tr(\partial_\mu \Sigma^\dagger 
\partial^\mu \Sigma)+c_2 h^{(v)}_\mu  h^{(v) *}_\nu  
 Tr(\partial^\mu \Sigma^\dagger 
\partial^\nu \Sigma+ \partial^\nu \Sigma^\dagger 
\partial^\mu \Sigma) \nonumber \\
&+&c_3 h^{(v)\mu}  h^{(v) *}_\mu  
Tr(v\cdot \partial  \Sigma^\dagger 
v \cdot \partial \Sigma), 
\end{eqnarray}
and similarly for the $2S$ state.
Here, and throughout the rest of this paper, we follow the notation
and normalizations in \cite {geo84}.
$\Sigma$ is a unitary $3\times 3$ matrix which contains Goldstone 
octet fields. $c_{1,2,3}$ are dimensionful parameters whose
scale at this point is unknown but will be determined in the next section.
$h^v_\mu$ is the spin one heavy quarkonia 
field, labeled by its velocity $v$, and satisfies
\beq
iv \cdot \partial h^{(v)}_\mu=0,~~v^\mu h^{(v)}_\mu=0.
\eeq
We will ignore terms which are off diagonal in the quarkonia
fields as they will give subleading contributions to the shift
in the level splitting. It is possible to make the spin symmetry
manifest by including the $\eta$ state along with the vector
$1S$ state as is done for the heavy-light system, but for
our purposes this is unnecessary.
The leading order chiral symmetry breaking piece of the Lagrangian is
\beq
L_{\chi sb}=c_4 h^{(v)\mu}  h^{(v) *}_\mu  
Tr(M (\Sigma +\Sigma^\dagger)),
\eeq
and $M$ is the diagonal quark mass matrix.  As will be shown below,
higher dimension terms in both the chirally symmetric as well as
chiral symmetry breaking pieces of the Lagrangian will be suppressed
by powers of $4\pi f$.

\subsection{Multipole and Twist Expansions}
\indent
It was pointed out by Gottfried \cite{got78} that interactions of long
wavelength gluons with quarkonia should be well described by a
multipole expansion, which yields an expansion in $E/(r_B^{-1})$ where
$E$ is the external gluonic energy scale and $r_B$ is the Bohr
radius. 
For decay processes this leads to an expansion in the relative quark
velocity  $v$. We are interested in the light quark mass dependence of
the mass splittings, and therefore the expansion parameter becomes
$m_\pi/(r_B^{-1})$. 
To implement the expansion we assume that
the decay goes through a two step process. Which is to say, the
hadronization process factorizes from the decay process. Thus, we
split the Hamiltonian up as follows
\beq
H=H_Q+H_g+H_{int},
\eeq
where $H_Q$ acts on the heavy quarks only and includes   the attractive
and repulsive Coulomb potentials for the singlet and octet states 
respectively. $H_g$  acts on the gluonic degrees of
freedom, and $H_{int}$ describes the interaction between the
quarkonium and the light degrees of freedom. We treat $H_{int}$
as a perturbation 
and the eigenstates of the leading order Hamiltonian are of
the factorized form
\beq
\mid \psi\rangle=\mid \phi\rangle \mid G\rangle.
\eeq
$\mid \!G\rangle$ corresponds to the state of the dynamical
gluons and is the ground state $\mid\! 0\rangle$ when the quarks
are in a relative color singlet state.  
The coupling to the spin will be higher order in a velocity
expansion.

The calculation of the matrix element for two gluon emission
then goes through much in the same way as the well known calculation
of Rayleigh scattering in QED.
Here, we shall simply state the result for the Euclidean space amplitude
and refer the reader to \cite{vol79}\cite{pes79} for details.
\beq
M=-\frac{g^2}{2N} \langle \phi_s \mid \vec{r}\cdot \vec{E}^a
\frac{1}{H_a-\epsilon-D^0} \vec{r}\cdot \vec{E}^a \mid \phi_s
\rangle,
\eeq
The wave function of the state in which the heavy quarks
are in a relative color singlet state, $\phi_s$, 
is an eigenstate of the Hamiltonian
containing the singlet part of the Coulomb potential and, 
$D_0$ acts on the gluonic Hilbert space. The factor
$1/(H_a-\epsilon-D^0)$ accounts for adjoint-state propagation 
between gluon emissions. 
$\epsilon$ is the eigenenergy of
$\phi$,  and $H_a$ is the repulsive adjoint Coulomb Hamiltonian acting
on the quark piece of the Hilbert space. 
This so called ``double-dipole'' amplitude is the leading
term in the multipole 
expansions.

The fact that the chromo-electric field $\vec{E}^a$ no
longer depends upon the relative separation of the quarks and
anti-quark allows us to write this amplitude in an OPE-like form where all the
details of the bound state are in the Wilson coefficient $\tilde{C}_n$. For S
wave states we may write
\beq
\label{twist1}
M=-\sum^\infty_{n=2~even} \tilde{C}_n \epsilon_0^{2-n} r_B^3
\left[\frac{1}{2}\vec{E}^a(D^0)^{n-2}\vec{E}^a\right],
\eeq
where
\beq
\label{cn}
\tilde{C}_n=\frac{16\pi}{N^2}\int \frac{d^3k}{(2\pi)^3}\frac{1}{3}
\left|\frac{\vec{r}}{r_B}\psi \right|^2\frac{1}{\left[
H_a/\epsilon_0+\epsilon/\epsilon_0\right]^{n-1}},
\eeq
\beq
r_B=\frac{16\pi}{g^2Nm_Q},~~\epsilon_0=\left(\frac{g^2N}{16\pi}\right)^2m_Q,
\eeq
and we have neglected spin symmetry breaking corrections.
We may therefore reproduce the interactions of long wavelength
gluons with the quarkonia via a (Minkowski) interaction Lagrangian
\beq
\label{eff}
L_{int}= \sum^\infty_{n=2~even} 
\tilde{C}_n{\epsilon_0^{2-n}}{r_B^3} h^{(v)\mu}  h^{(v) *}_\mu 
(\frac12\vec{E}\cdot(i D^0)^{n-2}\vec{E})+...
\eeq
where the ellipses denote higher multipoles as well as
higher orders in $\alpha_s$. This sum looks much like the leading
twist expansion in deep-inelastic scattering, except  the corrections
here are not power suppressed. Thus we will refer to  the expansion in the 
number of field insertions (read coupling expansion) as the twist
expansion\footnote{This similarity was utilized in \cite{pes79}.}.

If one assumes that the 
leading term in (\ref{eff}) dominates, then it is possible to
map this effective Lagrangian onto a chiral Lagrangian using symmetry 
arguments, as will be done
in the next section. However, there is no a priori reason why this should
be a good approximation. For decay processes, 
this expansion may  not be useful  
since the radiated gluon energies are the mass splittings (of the
order the Rydberg). Thus, as was pointed out in
\cite{lut93}, previous attempts at using this expansion to calculate
higher order corrections to quark mass relations \cite{don92}
are on dubious ground. However, there is phenomenological evidence to the 
effect 
that keeping only the leading term in (\ref{eff}) may not be unreasonable
for the case of decays (see e.g. \cite{vol80}). 
For the case of interest
here we are concerned with self interactions and
the OPE is a systematic expansion in $\Lambda_{\rm QCD}/\epsilon_0$.
We will come back to a detailed analysis regarding this issue 
in a future publication.

\section{Mapping Onto the Chiral Lagrangian}
  
It was pointed on in \cite{vol80} that if a given
double-dipole transition is dominated by the leading term in the 
OPE (\ref{twist1}), 
then 
it is possible
to map this interaction onto a chiral Lagrangian, thus allowing
us to 
reduce the number of unknown parameters.
Matrix elements of the double electric dipole operator $\alpha_s 
\vec{E}^a\cdot \vec{E}^a$, 
or the electric-magnetic dipole interference operator, $\alpha_s
\vec{E}^a \cdot \vec{H}^a$\cite{nov80},
can be mapped
to a chiral Lagrangian by using the fact that
\beq
\label{sym}
 \partial_\mu j_D ^\mu=T^\mu_\mu =
\frac{1}{2}\frac{\beta(g)}{g}G_{\mu \nu}^a 
G^{\mu \nu a}
+\sum_i(1-\gamma_i)\
m_i\overline{\Psi}_i\Psi_i.
\eeq
and
\beq
i\partial_\mu j^\mu_5=\frac{3\alpha_s}{8\pi}
G_{\mu \nu} ^a \tilde{G}^{\mu \nu a}+\sum_i
2m_i\bar{\Psi_i}\gamma_5\Psi_i.
\eeq
In the above equations, the sum is taken over the three light quark
flavors, $j_5^\mu$ is the axial current, $j_D^\mu$ is the dilatation
current and $T_{\mu \nu}$ is the stress energy tensor.

We see that knowing the symmetry transformation properties
of the Goldstone boson fields in the chiral Lagrangian
allows us to determine how the quarkonia couple to the light
degrees of freedom. This will allow us to disentangle the
many scales which arise in the effective field theory.
Furthermore,  both operators in (\ref{sym})
are renormalization group invariants\cite{inv}, and thus
 we need not worry about
at which scale to choose the value for the strong coupling.

Consider the case where the coupling to pions is 
dominated by the leading term in
the OPE of the 
double-dipole approximation, namely $\vec{ E}^2$. How does this operator
 map
into the chiral Lagrangian?
Following (\ref{sym}) we calculate the trace of the stress-energy
tensor for the chiral Lagrangian which is given
to order  $m_\pi^2$ in a chiral expansion
by
\beq
\partial_\mu j_D^\mu =T^\mu_\mu=-\frac{f^2}{2} 
{\rm Tr}(\partial_\mu \Sigma^\dagger \partial^\mu \Sigma)
-2f^2\mu {\rm Tr}(m\Sigma+m\Sigma^\dagger)+....
\eeq
Here the terms which are left off are down by powers of $4\pi f$.
Thus, given the analysis in the previous section, we now 
know that the lowest order coupling between the quarkonium state and
the light degrees of freedom is given by
\beq
\label{ans}
L_{int}=
-\frac{g}{2\beta(g)}\tilde{C}_2 r_B^3  h^{(v)*} \cdot h^{(v)} \left(\partial_\mu 
j_D^\mu-\sum_i(1-\gamma_i)\
m_i\bar{\Psi_i}\Psi_i\right).
\eeq
Here $\gamma_i$ is the anomalous dimension for the quark field $\Psi_i$.
We have neglected the term in the relation between $\vec{E}^a\cdot
\vec{E}^a$ and $G_{\mu \nu}G^{\mu \nu}$ which involves the magnetic
field as it is suppressed by spin symmetry.

We may now argue that the 
term coming form the anomalous dimensions
is small compared to one based on the idea that it has arisen from
integrating out shorter wavelengths and should thus be suppressed by
powers of the strong coupling evaluated at a perturbative scale.
If we drop this term then we are left with the following
coupling
\beq
\label{res}
L_{h^*h\pi}=\tilde{C}_2{r_B^{3}} \frac{g}{2\beta(g)} h^{(v)*} \cdot h^{(v)}
\left(f^2 {\rm Tr}\left(\partial_\mu \Sigma \right)
\left(\partial^\mu \Sigma^\dagger\right)+3f^2\mu {\rm Tr}
\left(m\Sigma+m\Sigma^\dagger\right)\right)+....
\eeq
If we are not willing to assume that the piece coming from the anomalous
dimension is small, then the relative sizes of the two terms in
(\ref{res}) becomes incalculable and we must include a new unknown
parameter into the Lagrangian. Notice that the assumption 
that the first term in the OPE of the leading twist sum dominates
effectively reduces the number of possible terms in the Lagrangian at
this order.
For instance the terms
\beq
f^2\bar{h}^v_\mu  h^v_\nu Tr \partial^\mu \Sigma 
\partial^\nu \Sigma^\dagger, ~
~\bar{h}_v \cdot h_v f^2Tr(v\cdot \partial) \Sigma 
(v\cdot \partial) \Sigma^\dagger
\eeq
are higher order in the OPE.

\subsection{Derivation of mass splittings formula}

We may now use (\ref{res}) to calculate the pion mass dependence
on the level splitting. The splitting will get a contribution from
a tree level counter-term which is independent of the pion mass, and
the leading dependence from the pion mass will come from
the second term in (\ref{res}).
At order $O(m_\pi^4)$ there will be
a one loop correction which will contribute a piece which is
non-analytic in the pion mass  as well as an unknown counter-term.
We ignore contributions which are 
 off diagonal in the quarkonia fields since they
contribute only at higher orders. The mass splitting is then
given by
\begin{eqnarray}
\label{answer}
\Delta m&=&A
+Bf^2\left[\frac{1}{(4\pi f)^2}
\left((m_\eta)^4\log \frac{m_\eta^2}{4\pi\mu^2}+3(m_\pi)^4
\log\frac{m_\pi^2}
{4\pi \mu^2}+4(m_K)^4\log\frac{m_K^2}{4\pi \mu^2}\right)
\right. \nonumber
\\  
&-&
\left.
6C((m_K)^2+\frac{(m_\pi)^2}{2})\right]
+\frac{D(\mu)}{16\pi^2}\left(3(m_\pi)^4+4(m_K)^4+(m_\eta)^4\right)
+.~.~.
\end{eqnarray}
The coefficients $A$, $B$, $C$ and $D$ are independent of the pion mass.
Any pion mass dependence that one might have expected to have arisen
from integrating out shorter wavelengths will
necessarily have a well defined Taylor expansion about $m_\pi=0$ and 
will contribute to a redefinition of $D$
and other higher order chiral symmetry breaking terms. 
The parameter $f$ is independent of the pion mass, and
for consistency we must know its value up to corrections
$O(m_\pi^4)$. At lowest order $f=f_\pi$ and corrections to
this relation can be calculated by computing the matrix element
of the axial current between the vacuum and the one pion state
as is usually done. There will be an unknown  counter-term which, 
in the three flavor theory, can be extracted by measuring the
ratio $\frac{f_\pi}{f_K}$\cite{gas85}. However, while $f$ is independent of
$m_\pi$, it is not independent of the number of flavors.
Thus, given that all the simulations to date were performed in the
two flavor theory, we can not determine the value of $f$ to
order $m_\pi^2$. Fortunately, this is unnecessary since all the
$m_\pi^2$ corrections will go into redefining the parameter $D$, 
which we are going to fit anyway. However, since we will not be
fitting the piece which is non-analytic in the pion mass we
use the relation \cite{lan73}
\beq
\label{f}
f=f_\pi(1+\frac{m_\pi^2}{(4\pi f_\pi)^2}\ln\frac{m_\pi^2}{4\pi\mu^2})+....
\eeq

Keeping only the leading term in the OPE, 
 we have in addition the constraints on the
coefficients:
\beq
C=1\eeq
and
\beq
\label{Bgiven}
B=\tilde{C}_2 r_B^3 \frac{2g}{\beta(g)}
\eeq
While $C=1$ is model independent, we see that the numerical value
of $B$ depends on the model wavefunction for the onium state.

If we reduce to the case of two families and use the relation
(\ref{f}) we arrive at
\beq
\Delta m =A_2-
B_2(2C_2m_\pi^2f_\pi^2+\frac{3m_\pi^4}{16\pi^2}
\ln\frac{m_\pi^2}{4\pi\mu^2})+\frac{D_2(\mu)}{16\pi^2}m_\pi^4.
\eeq
Again, when the first term in the OPE dominates, we have $C_2=1$ and 
$B_2=\tilde{C}_2 r_B^3
\frac{2g}{\beta(g)}$.

\section {Conclusions}

We have shown  that it is possible
to calculate the pion mass dependence of the levels splittings
in quarkonia up to a few unknown constants using a chiral
Lagrangian.
The number of unknown
constants which need to be extracted 
depends upon whether one is willing to accept the approximation that
the leading
term in the OPE of the leading multipole result dominates.
Within this approximation,  the number of unknown parameters can be reduced
by  mapping the dilatation  current in QCD onto the chiral
Lagrangian.

We have presented the a full result to order $m_\pi^4$, which could in
principle be used not only to calculate errors due to the pion mass,
but also errors due to $SU(3)$ breaking, should lattice calculations
reach the point where it is possible to calculated with three light
quarks.  Using the two flavor form of our results we have calculated
the errors in the extractions of $\alpha_{\overline{MS}}(M_Z)$ due to
the use of unphysical values of the pion mass. Our preliminary results
indicate that the errors in the extrapolation are of the same size as
the errors quoted in ref.~\cite{NRQCD1}. We emphasize that these
results are only preliminary, since we arrived at our numbers using
data points for which our expansion in $m_\pi^2$ is dubious.  We have
found however, that under the assumption that the unknown counter-term
at order $m^4_\pi$ is of the same order as the Log, the expansion is
well behaved.  Finally, it is very interesting to compare the scale
$f_\pi^2 r_B^3$ with the slope determined using the Monte-Carlo
data. We find that both numbers are on the order of $10^{-5}~MeV^{-1}$
which we find very encouraging.  This number is a factor of $10$
smaller than the value determined using Eq.~(\ref{Bgiven}) if the
model of ref.~\cite{pes79} is used for\footnote{While we do not trust
the model to predict $\tilde C_2$ accurately, we believe the sign,
which agrees with the one extracted from the Monte Carlo data, to be
trustworthy. } $\tilde C_2$.  Therefore, at worst we have
underestimated the error in the extrapolation.

\vskip1.75in

{\bf Acknowledgments.} We benefited from conversations with J. Shigemitsu, 
J. Kuti, A.
Manohar, C. Morningstar and S. Sharpe.
This work was supported in part by the Department of Energy under
Grant No.  DOE-FG03-90ER40546. B.G. was supported in part by a grant
from the Alfred P.~Sloan Foundation.

\end{document}